\documentclass[pra,aps,twocolumn,showpacs]{revtex4}
\usepackage{amsmath,graphicx,epsfig,color}
\begin{document}

\def\pd#1#2{\frac{\partial #1}{\partial #2}}

\title{\bf Interactions of vortices with rarefaction solitary
  waves in a Bose-Einstein condensate and their role in the
   decay of superfluid turbulence.}
\author {Natalia G. Berloff}
\affiliation {Department of Applied Mathematics and Theoretical Physics,
University of Cambridge, Wilberforce Road, Cambridge, CB3 0WA
}
\date {January 6, 2004}

\begin {abstract} There are several ways to create the vorticity-free
  solitary waves -- rarefaction pulses -- in condensates: by the process of strongly
  nonequilibrium condensate formation in a weakly interacting Bose
  gas, by creating local depletion of the condensate density by a
  laser beam, and by moving a small object with supercritical
  velocities. Perturbations created by  such
  waves colliding with vortices are studied in the context of the
  Gross-Pitaevskii model. We find that the effect of the interactions  consists of two competing mechanisms: the creation of vortex line as
   rarefaction waves acquire circulation 
  in a vicinity of a vortex core  and the loss of the vortex line to sound  due to
  Kelvin waves that are generated on vortex lines by rarefaction pulses.
  When a vortex ring
  collides with a rarefaction wave, the ring either stabilises to a smaller
  ring after emitting sound through Kelvin wave radiation or the entire
  energy of the vortex ring is lost to sound if the radius of the ring
  is of the order of the healing length. We show that during the time evolution of a 
  tangle of vortices,  the interactions with rarefaction
  pulses provide an important  dissipation mechanism enhancing  the
  decay of superfluid turbulence. 
  \end{abstract}
\pacs{ 03.75.Lm, 05.45.-a, 67.40.Vs, 67.57.De}
\maketitle
\section{Introduction}
Low temperature superfluids and recently discovered dilute
Bose-Einstein condensates are often modelled by the Gross-Pitaevskii
equation (GP), which is a nonlinear Schr\"oedinger equation on the
one-particle wave function, $\psi$. This equation provides a simple
framework for study many fundamental hydrodynamical properties of
condensates. In particular, vortex-sound interactions in condensates
are receiving increasing attention over last couple of years
\cite{vinen1, leadbeater1, br9, berloff, vinen2, leadbeater2}. It has
been suggested that the emission of sound by vortex reconnections and
vortex motion is the only active dissipation mechanism responsible for
the decay of superfluid turbulence. The decay of superfluid turbulence
via Kelvin wave radiation and vortex reconnections was studied in the
framework of the GP equation \cite{leadbeater2}. In this study the
collision of two small vortex rings were analysed and the loss  of the
vortex line due to  reconnections and 
Kelvin wave radiation was numerically evaluated.

The goal of this paper is to consider the effect of the vorticity-free
solitary waves that together with vortices are created in
Bose-Einstein condensates either during the self-evolution of a Bose
gas from a strongly nonequilibrium initial state or evolved from 
local density depletions of a condensate. We show that the
interactions with such waves  trigger the loss of the vortex line via Kelvin
wave radiation and that these interactions enhance the  dissipation of
the vortex tangle.

We write the GP equation in dimensionless form as
\begin{equation}
-2{\rm i} \pd \psi t =  \nabla^2 \psi +(1- |\psi|^2) \psi,
\label{gp}
\end{equation}
in dimensionless variables
such that the unit of length corresponds to the healing
length $\xi$, the speed of sound is $c=1/\sqrt{2}$,  and the density at
infinity is $\rho_\infty=|\psi_\infty|^2=1$.
The numerical integration was performed using a
   finite difference scheme. The faces of the computational box were
open to allow sound waves to escape; this is achieved numerically by
applying the Raymond-Kuo technique \cite{rk}. In the turbulence
simulations the periodic boundary conditions were used to conserve the
total number of particles in the system.

In a seminal paper, Jones
and Roberts \cite{jr} numerically  integrated the GP equation
(\ref{gp}) and determined the entire sequence
of solitary wave solutions of the GP equation, such as vortex rings,
vortex pairs, and finite amplitude sound waves named rarefaction
pulses. They showed  the location of the sequence on the
momentum, $p$,  energy, ${\cal E}$, plane.   In three dimensions they found two branches meeting at a cusp where $p$ and ${\cal E}$
assume their minimum values, $p_m$ and ${\cal E}_m$. As $p \to \infty$
on each branch, ${\cal E} \to \infty$. On the lower branch the
solutions are  asymptotic to  large vortex rings.

As ${\cal E}$ and $p$ decrease from infinity along the
lower branch, the solutions begin to lose their similarity to
large vortex rings.
Eventually, for a momentum $p_0$ slightly greater than $p_m$,
they lose their vorticity ($\psi$  loses its zero), and
thereafter the solitary solutions may better be described as
`rarefaction waves'. The upper branch consists entirely of these
and, as $p \to \infty$ on this branch, the solutions asymptotically
approach the rational soliton solution of the Kadomtsev-Petviashvili Type I
 equation. 

The Jones-Roberts  (JR) solitons are the only known disturbances that propagate with a
constant velocity. Notice, however, that there many other  waves that
change their velocity and shape during their motion. For instance, a
strong perturbation of a rarefaction pulse on the upper branch of the
dispersion curve causes it to collapse onto the lower branch and become
a vortex ring. During its transition the wave looses its energy and
momentum and the minimum of its density  decreases gradually before it 
reaches zero. In the next stage of the evolution the wave becomes a
non-axisymmetric ring  which radius increases until it reaches an
axisymmetric solitary state on the lower branch of JR dispersion
curve. All these intermediate time snapshots of the wavefunction of
the condensate have higher energy than any solitary wave on the lower
branch below the final axisymmetric vortex ring state although they
may have the same minima of the density in the transition. In what
follows we reserve the term ``rarefaction pulse'' or ``rarefaction
wave'' to describe a finite amplitude sound wave which is a JR
solitary wave.  The scenario just described illustrates
a typical mechanism in which the vortex line length may be created and increased as the
result of  perturbations of rarefaction pulses. In \cite{pade} we have
also considered a creation of vortex rings as a result of  energy and momentum
transfer between two interacting rarefaction pulses. But given a
complex tangle of interacting vortices we expect that the loss of
the vortex line length will dominate its  creation to account for the
experimentally observed decay of superfluid turbulence.

In \cite{pade} we have  developed an algorithm for finding
approximations to the JR solitary wave solutions with the correct asymptotic
behaviour at infinity. An axisymmetric solitary wave
moving with the velocity $U$ along the $x-$axis is accurately
approximated by  $\psi(x,s)=1+u(x,s) + {\rm i} \,v(x,s)$ where
\begin{eqnarray}
u&=&\frac{ a_{00}+a_{10}x^2 + a_{01}s^2+m c_{20}^{7/4}U(2x^2-f(U)s^2)}{(1+c_{10}x^2 +
  c_{01}s^2 + c_{20}(x^2+f(U)s^2)^2)^{7/4}},\nonumber\\
v&=&x\frac{b_{00}+b_{10}x^2 + b_{01}y^2-mc_{20}^{7/4}(x^2 + f(U)s^2)^2}{(1+c_{10}x^2 +
  c_{01}s^2 + c_{20}(x^2+f(U)s^2)^2)^{7/4}},
\label{pade}
\end{eqnarray}
where $a_{ij}, b_{ij}, c_{ij}$ and the dipole moment $m$ can be determined
from the series expansion and are functions of $U$. Also in (\ref{pade}), $s^2=y^2+z^2$, $x=x-Ut$,  and
$f(U)=1-2U^2$. Notice, that (\ref{pade}) represents a vortex ring as
well as a vorticity-free rarefaction pulse depending on $U$. If
$-2<a_{00}<-1$, then (\ref{pade}) represents a vortex ring as the
power series
expansion around zero shows; if $-1<a_{00}<0$, then (\ref{pade}) gives
an approximation of a rarefaction pulse, and  $a_{00}=-1$ is a
borderline case, such that the solitary wave has a single zero of the
wavefunction and, therefore, can be called a point defect.
These approximations can be used as initial
conditions in the numerical simulation that study interactions, when
the initial state is prepared by
multiplying the wavefunctions of the distant individual solitary waves.  Without an accurate starting point in
numerical calculations it would be impossible to separate clearly the effect
of interactions from the evolution of each  solitary wave  by itself as it
settled down from a poor initial guess. 

\section{Creation of rarefaction pulses in condensates}
Here we study the excitations created by  the collisions between
rarefaction pulses and vortices and the effect these collision have on the
evolution of vortex line length in a regime of superfluid 
turbulence with a tangle of vortices.  We start by describing three typical
scenario in which a large amount of rarefaction pulses is created, so
that their collisions with vortices do become significant.
The process of strongly nonequilibrium BEC formation in a
macroscopically large uniform weakly interacting Bose gas was
elucidated in \cite{svist} using numerical integration of the GP
equation. As the system evolves from  weakly turbulent state to state
of strong turbulence,  the phases of the complex amplitudes of the
wave field $\psi$ become strongly correlated and the period of their
oscillations become comparable with the evolution times of the
occupation numbers. This signifies that the  quasicondensate is formed with
appearance of a well-defined tangle of
quantized vortices and localized vorticity-free solitary
structures such as rarefaction pulses manifesting the start of the
final stage in the Bose gas evolution: the decay of superfluid
turbulence. The right panel of Fig. 5 presented in \cite{svist} shows a single vortex ring surrounded by many rarefaction
pulses as a result of the turbulent decay of the initial
vortex tangle, which also implies that many more rarefaction pulses were
present  at the beginning of the decay.

Next we describe how the JR solitons and in particular rarefaction pulses can be created in
Bose-Einstein condensates. It is generally believed that to create
vortices it is necessary to transmit angular momentum by rotationally
stirring the condensate with a laser beam. In \cite{bubble} we
demonstrated that  the collapse of a stationary spherically symmetric
  bubble can lead to the vortex nucleation. After the  condensate
  fills the cavity, it  begins to expand with growing 
density oscillations. These
'dips' of the density are themselves  unstable and the development of
this instability leads to the creation of localized disturbances:
vortex rings and rarefaction pulses.

Similarly, the evolution of a depletion of the condensate by laser
beam  after the laser was turned off may lead to creation of JR
solitons. Notice, that there is no need to add angular momentum or
deplete the density of the condensate to zero. We demonstrate this by
considering a depletion of the condensate, such that the wavefunction
of the condensate just before the laser is switched off is given by
\begin{equation}
\psi({\bf x})=\tfrac{1}{2}+\tfrac{1}{2}\tanh\Bigl[0.01(x^2+0.1(y^2+z^2)-100)\Bigr].
\label{ic}
\end{equation}
This density depletion roughly corresponds to the experiments in the
sodium condensates, see for instance \cite{raman}, with the healing length $\xi=0.3\mu m$
and the Gaussian beam waist of more than $10\xi$. The oblate
spheroidal form of the depletion (\ref{ic}) would be formed if such a
laser beam was guided in the condensate in a short straight streak of
the length of about $50\xi$.

We integrated the GP equation (\ref{gp}) using (\ref{ic}) as
our initial field in the box of the volume $V=120^3$. The faces of the computational box were
open to allow sound waves to escape. Fig. 1 shows the time snapshots of the density
cross-sections for $z=0$. Twelve vortex rings of various radii and four
rarefaction pulses moving outward from the center of the depletion are
clearly seen at $t=73$ snapshot. Fig.2 gives the isoplots of the
density at $|\psi|^2=0.3$.
Rarefaction pulses are seen as small oblate spheroids.

\begin{figure}[t]
\caption{(colour online) 
The snapshots of the contour plots of the density cross-section of a
condensate obtained by numerically integrating the GP model
(\ref{gp}) for the oblate spheroidal density depletion (\ref{ic}). 
Black solid lines show zeros of real and imaginary parts of
$\psi$, their intersection shows the position of
topological zeros.  Both low and high density regions are shown in
darker shades to emphasise  intermediate density regions. 
}
\centering
\bigskip
\bigskip
\epsfig{figure=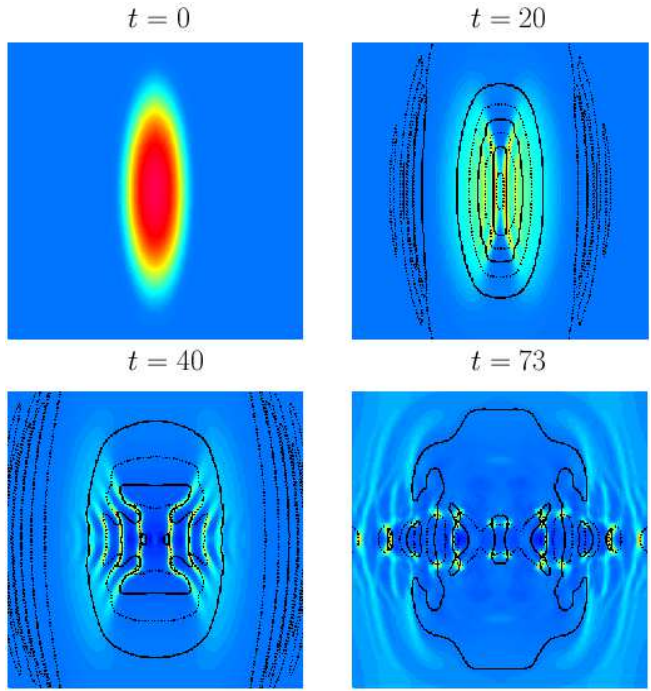, height = 3.6 in}
\end{figure}
\begin{figure}
\caption{(colour online) Time snapshots of the density isoplots
  $\rho=0.3$ of the evolution of the initial depletion of the
  condenstate amplitude given by (\ref{ic}).
}
\centering
\bigskip
\bigskip
\epsfig{figure=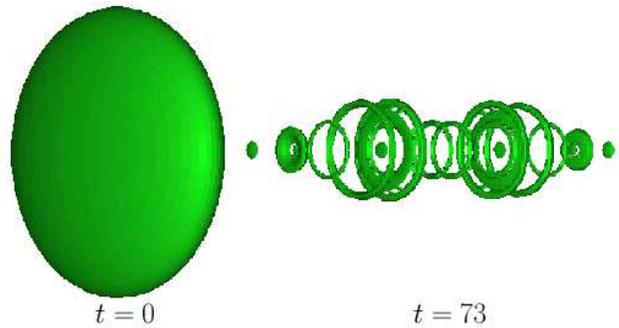, height =1.8in}
\end{figure}

The nucleation of vortices in a uniform
condensate  has been linked to
 critical velocities of the flow \cite{frisch, br7, br8}. It has also
 been pointed out \cite{rica} that a moving object of a size 
 comparable with the healing length generates rarefaction pulses rather
 than vortex rings when the velocity on the surface of this object
 exceeds the local speed of sound. Therefore, it is possible to
 generate the rarefaction pulses by guiding small objects with
 supercritical velocities through the condensate.

\section{Collisions with a straight line vortex}
In our first example we consider a collision between a rarefaction
pulse and the straight-line vortex of unit winding number. The rarefaction pulse is moving
with a constant self-induced velocity $U=0.63$ and belongs to the
lower branch of the JR dispersion curve. The sequence of density isosurface plots
illustrating the collision for two impact angles ($\pi/2$ and $\pi/4$)
is shown in Fig. 3. Initially the
center of the rarefaction pulse is 20 (Fig. 3(a)) or 10 (Fig. 3(b)) healing lengths apart from the axis of the
vortex. During the collision the rarefaction pulse creates a
distortion on the vortex line by exciting two Kelvin wave packets. In
the close vicinity of the vortex line, the rarefaction pulse acquired
vorticity it previously did not have and became a vortex ring
(Fig. 3(a) at $t=25$ and $t=30$ and Fig. 3(b) at $t=20$), so we would expect that the result of the
collision is similar to what was found in \cite{leadbeater2} for 
collisions of large vortex rings with small rings.
\begin{figure}
\caption{ Sequence of density isosurfaces for $\rho=0.3$
  illustrating a collision of the rarefaction pulse moving with the
  velocity $U=0.63$ and the straight line vortex. Two impact angles are
  shown: $\pi/2$ in (a) and $\pi/4$ in (b). The collision
  excites a pair of Kelvin wave packets propagating in the opposite
  directions along the vortex line.
}
\centering
\bigskip
\bigskip
\epsfig{figure=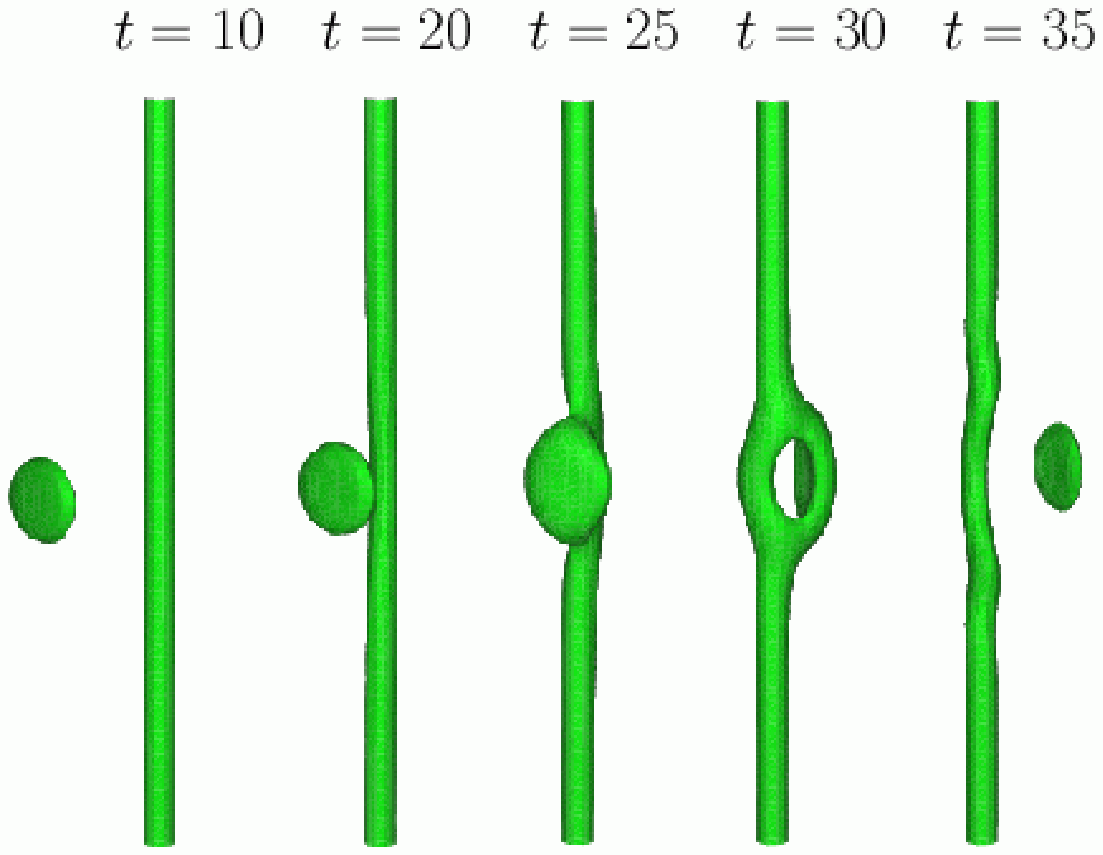, height = 2.in}
\epsfig{figure=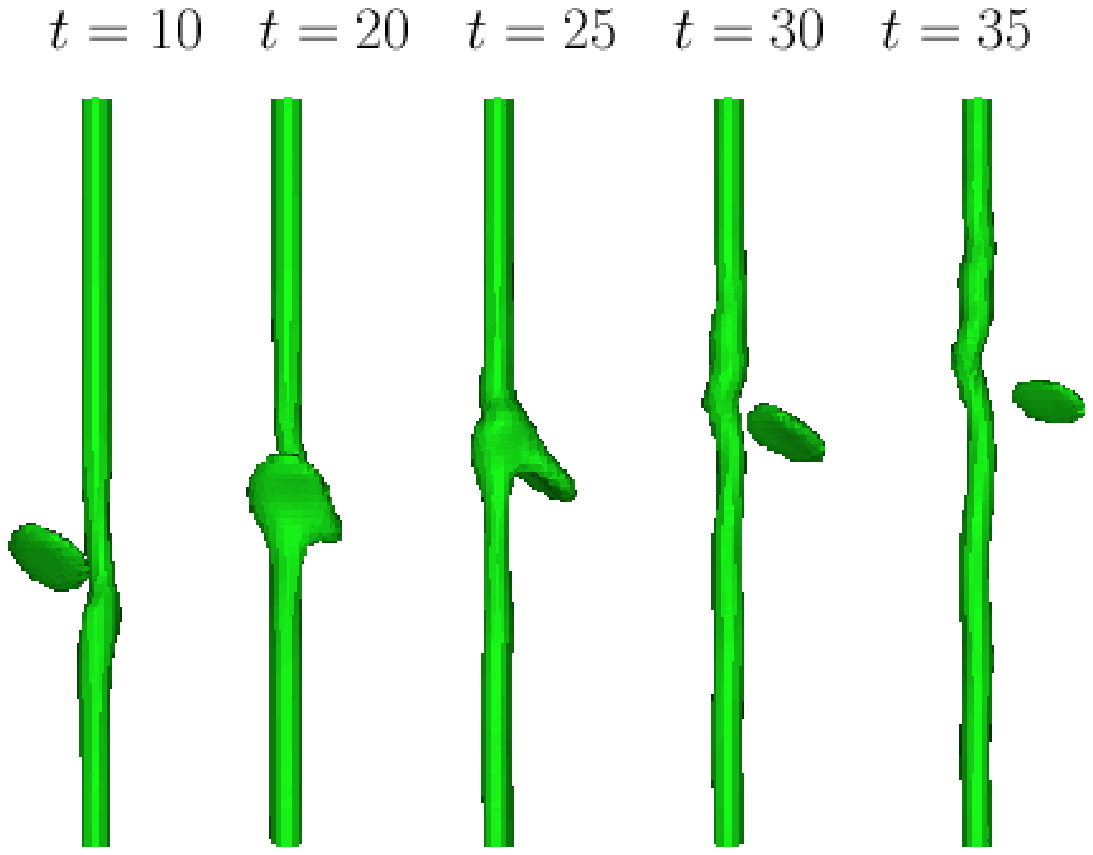, height = 2.in}
\begin{picture}(0,0)
\put(-200,250) {\Large (a)}
\put(-200,100) {\Large (b)}
\put(-86,210){\circle*{1}}
\end{picture}
\end{figure}

The dispersion relationship for Kelvin wave in the GP model in the
limit $k \rightarrow 0$ is  \cite{roberts02} 
\begin{equation}
\omega \sim \tfrac{1}{2}k^2(\ln(k) +0.003187...),
\label{disp}
\end{equation}
which gives a group velocity of the Kelvin wave packet as
\begin{equation}
U_g =-\frac{\pi}{\lambda}\Biggl(2 \ln[\frac{\lambda}{2\pi}]
-1.00637\Biggr),
\label{ug}
\end{equation}
where $\lambda$ is the dominating wavelength. Our calculations show
that $\lambda \sim 8$ healing lengths and $U_g \sim 0.2$ ($U_g\sim 0.3
c$), where $c$ is the speed of sound) which fits (\ref{ug}) quite
well. This result also agrees (when different scaling of the GP
equation is taken into account) with the wavelength of the excited
Kelvin wave during the vortex reconnections in \cite{leadbeater2}.  

The collision with a rarefaction pulse from the upper branch of the JR
dispersion 
curve excites a Kelvin wave packet of a similar central
wavelength. Fig. 4 shows the contour plots of the density
cross-sections of two such collisions for
two different rarefaction pulses that belong to lower (left panel) and
upper (right panel) branches of the JR dispersion curve. The wavelength
of the created excitation is best seen by the intersections of zeros
of real and imaginary parts of the wavefunction $\psi$ shown by solid and dashed lines
correspondingly. Similarly,  a collision with an offset generates a
Kelvin wave packet of approximately the same dominating wavelength but
a smaller amplitude. 
\begin{figure}
\caption{(colour online) The snapshots of the 
contour plots of the density cross-section of a condensate obtained by numerically
integrating the GP model (\ref{gp}) illustrating the collision of the rarefaction
pulses and the straight line vortex at $t=35$ that are initially a distance of 20
healing lengths apart. Left panel shows the collision of the
rarefaction pulse from the lower branch of the JR dispersion curve ($U=0.63, {\cal E}
=52.3, p=72.2, m=8.37$) and the right panel is for the
rarefaction pulse from the upper branch of the JR dispersion curve ($U=0.68, {\cal E}
=53.7, p=74.1, m=8.8$). Black solid and dashed lines show zeros of real and
imaginary parts of $\psi$ correspondingly, therefore, their intersection shows the
position of topological zeros.  Both low and high density regions are
shown in 
darker shades to emphasise  intermediate density regions. Only a portion of an actual computational box is shown.}
\centering
\bigskip
\bigskip
\epsfig{figure=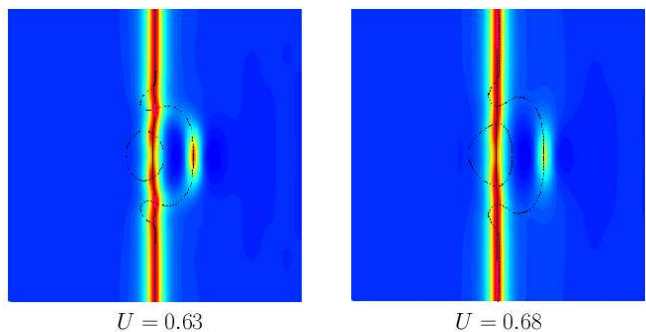, height = 1.7in}
\end{figure}

We can estimate the amplitude of the Kelvin wave generated by a
{\it distant }
rarefaction wave by reducing the problem to two dimensions. We shall
assume that initially the straight line vortex is  positioned at
$(0,0,z)$ and the rarefaction wave is moving in the positive $x-$direction
along $y=-d_y < 0$ with $\xi \ll d_y$. Under these assumptions, the wavelength of the
perturbation along the vortex line is much larger than its amplitude,
$\eta$, 
(the maximum displacement of the vortex center from the origin in
$xy-$plane), therefore, we can assume that the $x-$ and $y-$
components of the force exerted by a rarefaction wave on the vortex dominate the
$z-$component (in other words, the vortex may be considered as rigid). Also,
if the incident angle of the rarefaction pulse and the direction of
the vortex axis is nonzero, the amplitude calculated below would have to be multiplied
by the cosine of this angle. In $xy-$plane the vortex is advected by the rarefaction
pulse, so that it instantaneous velocity coincides, to the leading
order, with the velocity of the rarefaction pulse at the vortex
location. The center of the rarefaction pulse is moving with the  self-induced velocity, ${\bf U} =
(U,0,0)$ and is advected
by the vortex. The motion of the vortex and of the rarefaction pulse is,
therefore, given by the system of the coupled ordinary differential
equations
\begin{eqnarray}
\frac{d x^r}{dt}&=&U - \frac{y^r-y^v}{(x^r-x^v)^2+(y^r-y^v)^2},\nonumber\\
\frac{d y^r}{dt}&=& \frac{x^r-x^v}{(x^r-x^v)^2+(y^r-y^v)^2},\\
\frac{d x^v}{dt}&=&V_x(x^v-x^r,y^v-y^r),\nonumber\\
\frac{d y^v}{dt}&=&V_y(x^v-x^r,y^v-y^r),\nonumber\\ \nonumber
\label{coupled}
\end{eqnarray}
where $(x^r(t),y^r(t))$ and $(x^v(t),y^v(t))$ are the positions of the centres of the
rarefaction pulse and the vortex, correspondingly, at time $t$. $V_x$
and $V_y$ are
the velocity components induced by the rarefaction pulse on the vortex
given by
$V_x=\partial S/\partial x$ and $V_y=\partial S/\partial y$, where $S$
is the phase of the wavefunction $\psi=R \exp^{i S}$, so that 
\begin{equation}
dS=\frac{(1+u)u\,dv-v \,du}{(1+u)^2+v^2},
\label{ds}
\end{equation}
where the real, $1+u$, and imaginary, $v$, parts of $\psi$ are
approximated by (\ref{pade}). The initial conditions are given by
$(x^v(0),y^v(0))=(0,0)$ and $(x^r(0),y^r(0))=-{\bf d}=-(d_x,d_y)$. Fig. 5 shows
the trajectories of the vortex for $U=0.63$ and ${\bf d}=(10,10), (10,20), (20,10)$
and $(20,20)$. It is clear from the figure, that the maximal amplitude
of the displacement, $\eta$, does not depend strongly on $d_x$, but on
$d_y$. Fig. 6 plots  $\eta$ as a
function of $d_y$ for two rarefaction pulses that belong to the lower ($U=0.63$) and
upper ($U=0.69$) branches of the JR dispersion curve. For large
offsets, $d_y$, ($d_y
> 30$) the maximum
amplitude, $\eta$, decays exponentially with $d_y$, so that $\eta \sim
d_y^{1.24}$ for $U=0.63$ and $\eta \sim
d_y^{1.62}$ for $U=0.69$. 
\begin{figure}
\caption{(colour online)  The parametric plots of the trajectories
  $(x^v(t), y^v(t))$ of the vortex initially at $(0,0)$ and moving in $xy-$plane under the
  influence of the rarefaction pulse initially at $(-20,-20)$,
  $(-20,-10)$, $(-10,-20)$ and $(-10,10)$. The vortex travelled the
  distance between two adjacent points in unit time. }
\centering
\bigskip
\bigskip
\epsfig{figure=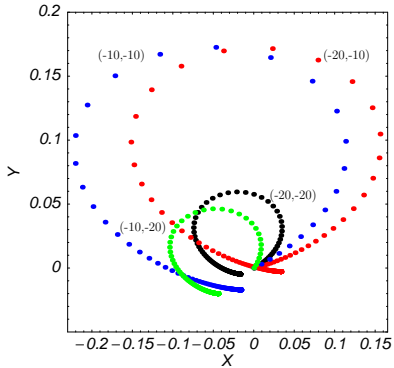, height = 2.in}
\end{figure}
\begin{figure}
\caption{The amplitude of the maximum displacement of
  the vortex from the origin as the result of  the interactions with
  the rarefaction pulse from the lower ($U=0.63$) and upper ($U=0.69$)
  branches of the JR dispersion
  curve as functions of the initial distance $d_y$ between the vortex
  and the center of the rarefaction pulse along the $y-$axis.}
\centering
\bigskip
\bigskip
\epsfig{figure=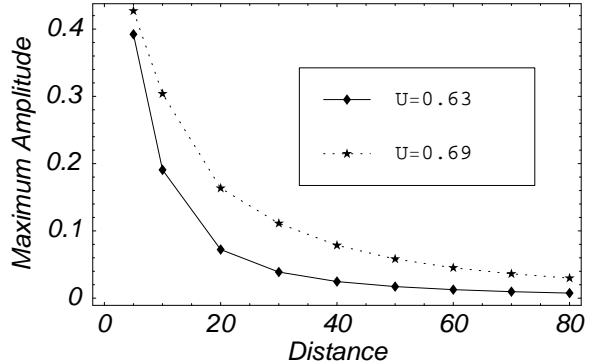, height = 2in}
\end{figure}
\section{Collisions with a vortex ring}
Next we consider the collision of a rarefaction pulse with a vortex
ring. As in the previous cases, 
such a  collision excites  Kelvin wave with release of sound
energy. Fig. 7 illustrates the time snapshots of the collision between
the vortex ring of radius $R=5.1$ moving along the $x-$axis  and the
rarefaction pulse moving along the line $y=-R$. The total vortex line
is dramatically increased at time $t\sim 11$ (from the line length
${\ell} \sim 32$ to ${\ell} \sim 50$) due to the rarefaction
pulse being transformed into a small vortex ring during a short time
interval, $\delta t\sim 5$. After that the ring loses its energy to
Kelvin wave radiation and stabilises to axisymmetric vortex ring of slightly
smaller radius at $t\sim 100$, so that the total loss of the vortex
line is about $3\%$.  The length of the vortex line does not
change for $t > 100$.  We repeated the calculations for the rarefaction
pulse from the upper branch of the JR dispersion curve and found that
the ring stabilises after losing $4\%$ of its length. These results may seem to contradict the findings
of \cite{leadbeater2}, where the length of the vortex line continues
to decay after the collision. The calculations of \cite{leadbeater2}
were done in computational box with periodic boundaries, so the sound
waves generated during the reconnections of a large ring with a
small ring continue to excite Kelvin waves on the vortex line with a
continuous loss of the vortex line to sound radiation. The rate of the
decay of the vortex line in this case should depend on the size of the
computational box. In our
calculations the faces of the computational box were
open to allow sound waves to escape \cite{rk}, which allowed us to
compute the effect of a single collision with the finite amplitude sound wave.

For very small vortex rings ($R\sim 1$) the collision with a
rarefaction pulse causes all the energy and momentum to be  converted
into sound waves with a complete loss of the vortex line.
\begin{figure}
\caption{(colour online) The time snapshots of the density isosurfaces
  of the condensate 
  at $|\psi|^2=0.3$ for the collision of the vortex ring of radius
  $5.1$ and the rarefaction pulse moving with the velocity $U=0.63$. The density contour plots at the cross-section
  for $z=0$ are given below each isosurface. Black solid and dashed lines show zeros of real and
imaginary parts of $\psi$ correspondingly, therefore, their intersection shows the
position of topological zeros.  Both low and high density regions are
shown in 
darker shades to emphasise  intermediate density regions. Only a portion of an actual computational box is shown. }
\centering
\bigskip
\bigskip
\epsfig{figure=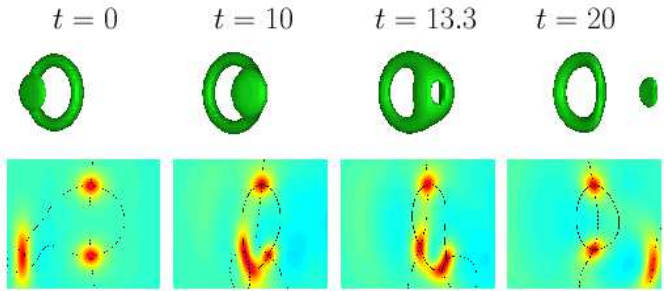, height = 1.55in} 
\end{figure}
\section{Decay of a vortex tangle}
Next, we  estimate the effect that the interactions with
rarefaction pulses have on a vortex tangle. This time we 
perform the simulations similar to those of
\cite{kivotides,leadbeater2}, in which the vortex
tangle was created by four colliding rings in a periodic box. But  in
our simulations we  regulate the initial amount of sound waves in our
computational box by
introducing them as randomly oriented rarefaction waves in addition to
the vortex rings.

For simplicity we shall assume that  all 
sound waves initially consist of three types of rarefaction pulses
moving in random directions. The approximations to these solitary
waves, that were developed in \cite{pade}, become very useful in the construction of our initial
conditions. If the rarefaction pulse given by (\ref{pade}) is rotated by angle $\alpha$ to the $x-$axis in the
$(x,y)$ plane and  by angle $\beta$ to the $y-$axis in the $(y,z)$
plane, then its wavefunction is  given by $\psi(x',s')$ where
$x'=x \cos\alpha - y \sin\alpha$, $s'=\sqrt{(y'\cos\beta
  -z\sin\beta)^2+(z\cos\beta + y'\sin\beta)^2}$, and $y'=y\cos\alpha +
  x\sin\alpha$. The angles of rotation $\alpha$ and $\beta$ as well as the
  center of the rarefaction pulse  are
   chosen randomly.

\begin{figure}
\caption{(color online) The vortex line length for the collision of four vortex rings
  of radius $R=30.1$. Initially the number of rarefaction pulses is
  $N_{rare} = 0$ (black line), $N_{rare}=100$ (red or dark grey line)
  and $N_{rare}=200$ (green or light grey line). A decay characterized
  by $\chi_2 = 0.3$ is plotted (thick line) to show the experimentally
  determined decay of the vortex line length. 
 }
\centering
\epsfig{figure=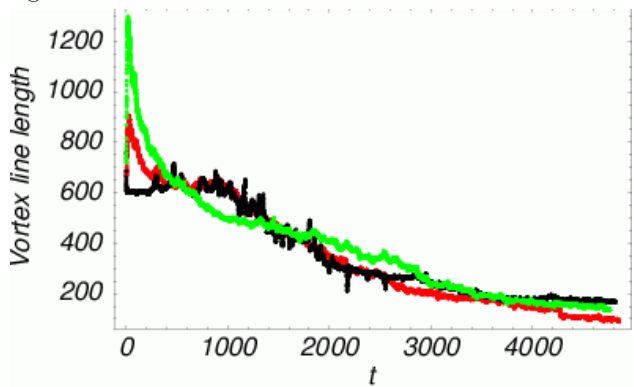, height = 2.in}
\end{figure}
\begin{figure}
\caption{Density isosurfaces ($|\psi|^2=0.3$) showing the  snapshots
  of the time evolution of the initial state consisting of four vortex
  rings of radius $R=30.1$ and $100$ rarefaction pulses. Many small
  vortex rings are formed during the evolution. 
}
\centering
\epsfig{figure=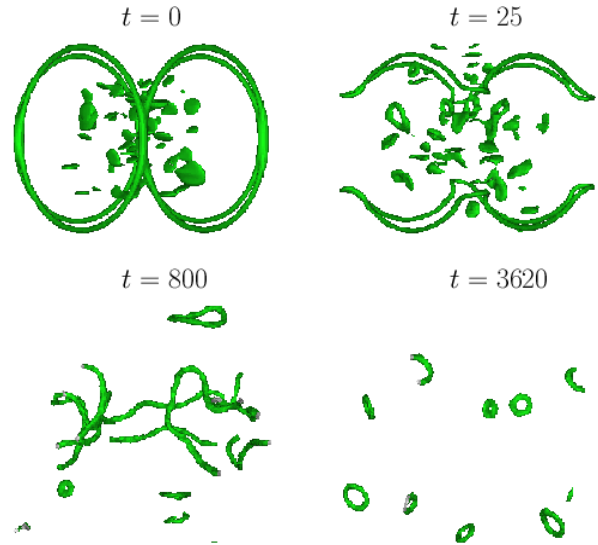, height = 3.in}
\end{figure}

Our calculations were performed in a periodic box with volume,
$V=80^3$, which is divided by $161^3$ grid points with a spacing
$0.5$. The numerical scheme is 4th order globally accurate with 4th
order Runge-Kutta time integration with the time step $0.025$. A grid spacing of $0.25$
together with the 
time step $0.00625$ was also used to test the accuracy of the
numerical method. Initially there are four vortex rings of the radius
$R=30.1$ centered at $(R,1,0), (-R,-1,0), (1, -R,0)$ and $(-1,R,0)$
that are  moving towards the center of the box. $N_{rare}$ rarefaction pulses are
distributed randomly in the internal half of the box away from the
vortices. The results for the dynamics of the vortex line length for three
sets of computations with $N_{rare}=0, 100$ and $200$ are plotted in
  Fig. 8. The density isosurfaces plots for $N_{rare}=100$ are given
  in Fig. 9. In the case of nonzero number of rarefaction pulses there
  is a rapid initial growth of the length of the vortex line due to the
  evolution of some of the rarefaction pulses into small vortex rings
  as the result of 
  the energy transfer among them \cite{pade}. Consequently, the length of
  the vortex line shows the balance between vortices being created and
  vortex line being destroyed due  to reconnections and Kelvin wave
  radiation. Our simulations show that apart from the initial growth
  of the 
  vortex line in the simulations with $N_{rare}=100$ initially, the dynamics for $N_{rare}=100$
  and 
$N_{rare}=0$ is very similar for in the time interval $[400,2600]$. On the other hand, the
  case with $N_{rare}=200$ is quite different. During the time
  interval $[500, 1300]$, the decay of the vortex tangle 
  is much faster than for two other runs. Whereas after
  $t\sim 1800$ it is the run with $N_{rare}=200$ that shows a slower
  decay. We speculate that this can be best explained by the
  following. During the moderate times, the vortex line loss due to the excitations of the vortex
  line created by rarefaction pulses depends on how many rarefaction
  pulses survived as their energy is gradually being converted to
  high-frequency waves. The more of rarefaction pulses were present in the initial
  state, the longer at least some of them are present in the system,
  therefore, the state
  with $N_{rare}=200$ shows a faster decay during the moderate times. At
  larger times, all of the rarefaction pulses are destroyed by the
  collisions and vortices live on the background of high-frequency
  waves that limit their mobility  and reduce the number of
  reconnections which slows down the decay of
  the vortex line length, with the total energy carried by these
  high-frequency waves being higher for  the simulations with
  $N_{rare}=200$ initially. 

For even larger times the situation is reversed one more time as
reconnections are now infrequent and the main decay mechanism is
through Kelvin wave radiation. Therefore, the rate of decay of vortex
line is propotional to the amount of high-frequency waves (slowest
decay for $N_{rare}=0$, fastest decay for $N_{rare}=200$.)

 For Kelvin-wave cascade, where energy is transfered to a much shorter
 wavelengths with a cutoff below a critical wavelength, the vortex
 line density $L=\ell/V$ can be described by the Vinen equation \cite{vinen}
\begin{equation}
\frac{d L}{d t} = -\frac{\kappa}{2 \pi}\chi_2 L^2,
\label{L}
\end{equation}
where  $\kappa=2\pi$ in our
dimensionless units and  $\chi_2$ is a
dimensionless coefficient. More accurately $\chi_2$ is a  weak
(logarithmic) function of $L$ and of other parameters such as the
Kelvin cutoff and temperature.  The logarithmic dependence on $L$ can
be easily obtained in the context of the local induction
approximation. The presence of a large amount of rarefaction pulses
changes this weak dependence on $L$ and the decay of the vortex line
approaches an exponential decay instead:
\begin{equation}
\frac{d L}{d t} = -\frac{\kappa}{2 \pi}\chi_2 L.
\label{Lexp}
\end{equation}
In Fig. 10 we plotted the curve (the solution
of (\ref{Lexp}))
\begin{equation}
\ell = \ell_0 \exp(-\chi_2 t)
\label{l0}
\end{equation}
 for $
\chi_2=0.00035$ and $\ell_0=760$ to illustrate the exponential decay
of the vortex line length.
\begin{figure}
\caption{(color online) The vortex line length for the collision of four vortex rings
  of radius $R=30.1$. Initially the number of rarefaction pulses is
  $N_{rare}=200$ (green or light grey line). The black solid line gives
  a decay of the vortex line given by (\ref{l0}) for $\ell_0=760$ and $\chi_2=0.00035$. }
\centering
\epsfig{figure=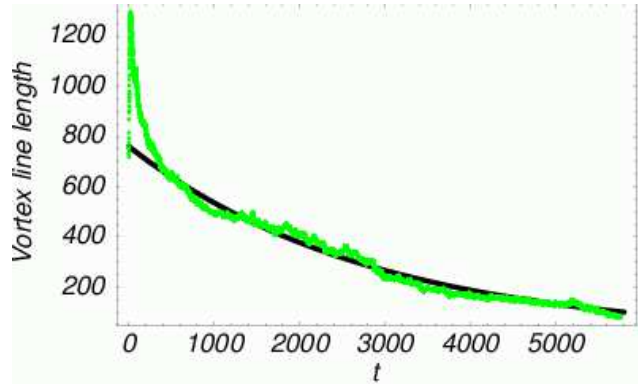, height = 2.in}
\end{figure}

Finally, we note that the presence of rarefaction waves in a
condensate with  regular vortex lattices has destabilising effect and
facilitates reconnections. If
vortices are antiparallel, the Kelvin wave created on the filaments as
the result of interaction with rarefaction waves facilitates the
growth of the Crow instability \cite{br9} that leads to vortex
reconnections that destroy a significant amount of vortex line (as we
 estimated in  \cite{br9}, the minimum fractional line loss of the
 pair of antiparallel vortices that are initially the distance $h$
 apart and  perturbed by a wave of the wavenumber $k \ll 1$ is about
 $1-\sqrt{kh/2}$). The interactions of the rarefaction waves with the
 lattice of parallel vortices (created, for instance, in rotating
 condensates) also destabilises the array and creates a tangle as
 Fig. 11 illustrates.

\begin{figure}
\caption{Density isosurfaces for $|\psi|^2=0.3$ showing two snapshots
  of the time evolution of the array of $10$ initially straight parallel
  vortices that interacted with $10$ rarefaction pulses. 
}
\centering
\bigskip
\bigskip
\epsfig{figure=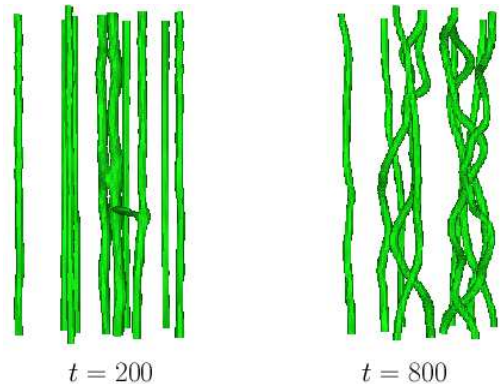, height = 2.in}
\end{figure}
\section{Conclusions}

In summary, we established several mechanisms for creation of the
rarefaction pulses in condensates. We have studied the effects of their interactions  with straight line
vortices, vortex rings and tangles of vortices. We showed that there
are two competing mechanisms of these interactions. Firstly, the
interactions of waves at close distances lead to the increase of the
vortex line as the rarefaction pulses in the regions of lower density (in the vicinity of the vortex core
or each other) may acquire circulation and become vortex
rings. Secondly, rarefaction waves excite  Kelvin waves on the vortex
filaments causing a loss of the vortex line due to sound
emission. At high vortex line densities our simulations suggest that
the Kelvin wave radiation is enhanced by the presence of rarefaction
pulses and this accounts for a dramatic increase in the rate of the decay of the
vortex line length.
\acknowledgments

The author acknowledges the support from the NSF grant
DMS-0104288. The author is very greateful to Professor Paul Roberts for many
discussions about the structure of rarefaction pulses and the importance
of sound emission during her tenure at UCLA. The author is indebted to
Professor Boris Svistunov for many useful comments about this manuscript.


\begin{thebibliography}{99}

\bibitem{vinen1}
W.F. Vinen, Phys. Rev. B {\bf 64}, 134520 (2001).

\bibitem{leadbeater1}
M. Leadbeater, T. Winiecki, D.C. Samuels, C.F. Barenghi and C.S. Adams,
Phys. Rev. Letters {\bf 86} 1410 (2001).

\bibitem{br9}
N.G. Berloff and P.H. Roberts, J. Phys. A: Math. Gen., {\bf 34}
10057 (2001)

\bibitem{berloff} 
N.G. Berloff,  Phys. Rev. B, {\bf 65} 174518 (2002)


\bibitem{vinen2}
W.F. Vinen, M. Tsubota and A. Mitani,
Phys. Rev. Letters {\bf 91}, 135301 (2003).

\bibitem{leadbeater2}
M. Leadbeater, D.C. Samuels, C.F. Barenghi and C.S. Adams,
Phys. Rev. A {\bf 67} 015601 (2003).




\bibitem{rk}
G.W. Raymond and H.L. Kuo,
  Q. J. R. Meteorol. Soc. {\bf 110}, 525 (1984)

\bibitem{jr} C. A. Jones and P. H. Roberts, J. Phys.\ A: Gen.\ Phys.
{\bf 15},  2599 (1982).


\bibitem{pade}
N. G. Berloff J. Phys.: Math. Gen. {\bf 37}, 1617 (2004)

\bibitem{svist}
N.G. Berloff and B. V. Svistunov, Phys. Rev. A, {\bf 66}, 013603 (2002)

\bibitem{bubble} 
N.G. Berloff and C.F.Barenghi, submitted to Phys. Rev. Lett,
cond-mat/0401021.

\bibitem{raman} C. Raman, M. K\"ohl, R. Onofrio et al.,
Phys. Rev. Lett, {\bf 83}, 2502 (1999)


\bibitem{frisch}
T. Frisch, Y. Pomeau and S. Rica, Phys. Rev. Lett., 
{\bf 69}, 1644 (1992) 

\bibitem{br7}
N. G. Berloff and P. H. Roberts,
 J. Phys.: Math. Gen. {\bf 33}, 4025 (2000)

\bibitem{br8}
N.G. Berloff and P.H. Roberts, J. Phys. A: Math. Gen., {\bf 34}
81 (2001)

\bibitem{rica} Sergio Rica, private communications.
\bibitem{roberts02} P. H. Roberts, Proc. R. Soc. Lond. A, {\bf 459},
  597 (2002)

\bibitem{kivotides}
D. Kivotides, J.C. Vassilicos, D.C. Samuels and C.F. Barenghi,
Phys. Rev. Letts., {\bf 86}, 3080 (2001).


\bibitem{vinen}
W.F. Vinen, Proc. R. Soc. London. Ser. A {\bf 242}, 493 (1957)

\end{thebibliography}
\end{document}